\documentclass[namedreferences]{SolarPhysics}
\usepackage{natbib}                     	% For citations: redefine \cite commands
\usepackage[optionalrh]{spr-sola-addons} 	% For Solar Physics
\usepackage{solaheader}				% FOR PREPRINT SERVERS
\usepackage{graphicx}
\usepackage{rotating}
\usepackage{courier}  				% Change the \texttt command to courier style
\usepackage{amssymb}                    	% useful mathematical symbols
\usepackage{color}                       	% For color text: \color command
\usepackage{url}                         	% For breaking URLs easily trough lines
\newcommand{\etal}{ et al. }
\newcommand{\be}{\begin{equation}}
\newcommand{\ee}{\end{equation}}
\newcommand{\beq}{\begin{eqnarray}}
\newcommand{\eeq}{\end{eqnarray}}
\newcommand{\bl}{\color{black}}
\newcommand{\adv}{    {\it Adv. Space Res.}}

\newcommand{\aap}{    {\it Astron. Astrophys.}}
\newcommand{\araa}{   {\it Ann. Rev. Astron. Astrophys.}}
\newcommand{\aaps}{   {\it Astron. Astrophys. Suppl.}}
\newcommand{\aapr}{   {\it Astron. Astrophys. Rev.}}
\newcommand{\ag}{     {\it Ann. Geophys.}}

\newcommand{\apj}{    {\it Astrophys. J.}}
\newcommand{\apjl}{   {\it Astrophys. J. Lett.}}

\newcommand{\grl}{    {\it Geophys. Res. Lett.}}

\newcommand{\jgr}{    {\it J. Geophys. Res.}}

\newcommand{\solphys}{{\it Solar Phys.}}

\newcommand{\ssr}{    {\it Space Sci. Rev.}}

\newcommand{\RSUN}{R$_{\odot}$}
\newcommand{\Ha}{H$\alpha$}
\begin{document}
\begin{article}
\begin{opening}
\title{The 17 January 2005 Complex Solar Radio Event Associated with Interacting Fast Coronal Mass Ejections }
\author{A. \surname{Hillaris}$^{1}$,
	O. \surname{Malandraki}$^{2}$,
	K.-L \surname{Klein}$^3$,
	P. \surname{Preka-Papadema}$^{1}$,
	X. \surname{Moussas}$^{1}$,
	C. \surname{Bouratzis}$^{1}$,
	E. \surname{Mitsakou}$^{1}$,
	P. \surname{Tsitsipis}$^{4}$,
	A. \surname{Kontogeorgos}$^{4}$}
%-------------------------------------------------
\runningauthor{Hillaris \etal}
\runningtitle{Radio Observations of the 17 January 2005 CME--CME Interaction}
\institute{$^{1}$ Section of Astrophysics, Astronomy and Mechanics, Department of Physics,
		University of Athens, Zografos (Athens) , GR-15783, Greece
		email: \url{xmoussas@phys.uoa.gr} email: \url{kbouratz@phys.uoa.gr} 
		email: \url{ppreka@phys.uoa.gr}  email: \url{ahilaris@phys.uoa.gr} 
		email: \url{emitsaku@phys.uoa.gr} \\
		$^{2}$ Institute of Astronomy and Astrophysics, National Observatory of Athens,
		11810--Athens, Greece email: \url{omaland@astro.noa.gr}\\         
		$^{3}$ Observatoire de Paris, LESIA-CNRS UMR 8109, Univ. Paris 6 \& Paris 7, 
                                Observatoire de Meudon, F-92195 Meudon, France
		email: \url{ludwig.klein@obspm.fr}\\
		$^{4}$ Technological Educational Institute of Lamia, GR-35100 Lamia, Greece
		email: \url{tsitsipis@teilam.gr} email: \url{akontog@teilam.gr}\\}
%-------------------------------------------------
\begin{abstract} 
{On 17 January 2005 two fast coronal mass ejections were recorded in close succession during two distinct 
episodes of  a 3B/X3.8 flare. Both were accompanied by metre-to-kilometre type-III 
groups tracing energetic electrons  that escape into the interplanetary space and by 
decametre-to-hectometre type-II bursts attributed to CME-driven shock waves. {\bl A peculiar type-III burst group 
was observed below 600 kHz 1.5 hours after the second type III group. It occurred without any simultaneous activity  at %KLK 
higher frequencies, around the time when the two CMEs were expected to interact.}
We associate this emission with the interaction of the CMEs at heliocentric distances of about 25~\RSUN.
Near-relativistic electrons observed by the  EPAM experiment onboard ACE near 1~AU revealed successive particle 
releases that can be associated with the two flare/CME events and the low-frequency type-III burst at  the time 
of CME interaction. We compare the pros and cons of shock acceleration and acceleration in the course of 
magnetic reconnection for the escaping electron beams revealed by the type III bursts and 
 for the electrons measured {\it in situ}.}
\end{abstract}
%-------------------------------------------------
\end{opening}
%-------------------------------------------------
\section{Introduction}\label{Introduction}

The acceleration of charged particles to high energies in the solar corona is related to flares, which reveal 
the dissipation of magnetically stored energy in complex magnetic field structures of the low corona, and 
to coronal  mass ejections (CMEs), which are large-scale, complex magnetic-field-plasma structures ejected from the Sun. 
CMEs can drive bow shocks, and their perturbation of the coronal magnetic field can also give rise to magnetic 
reconnection, where energy can be released in a similar way as during flares. 

When several  CMEs are launched along the same path, a faster CME may overtake 
a slower preceding one, and the two CMEs can merge into a single structure.
For this phenomenon \citet{Gopalswamy01} 
introduced the term {\em{CME Cannibalism}}.  The CME-CME interaction 
was found associated with a characteristic low-frequency  continuum
radio emission. \citet{Gopalswamy02}  interpreted this type of activity as the radio signature of 
non-thermal electrons originating either during reconnection between the two CMEs or as the shock of 
the second, faster CME travels through the body of the first \citep[see ][for a numerical study of 
two interacting coronal mass ejections]{Schmidt04}.

In this paper we use radio diagnostics to study electron acceleration during a complex 
solar event broadly consisting of two stages, {\bl each associated with a 
distinct episode of a flare} and with a fast CME, which occurred in close temporal succession on 17 January 2005. The CMEs interacted 
at a few tens of {\RSUN}. Both the flare/CME events and the CME interaction were accompanied by radio emission, 
which is used here to study electron acceleration scenarios. Energetic electrons in the corona and interplanetary 
space are traced by their dm-to-km-wave radio emission, mostly excited at or near the electron plasma frequency. 
The emission provides a diagnostic of the type of the exciter and its path from the low corona (cm-dm wavelengths) 
to 1~AU (km wavelengths). 
%%--------------------------------------------------------------------------------------------------
Radio emissions from exciters moving through the corona appear in dynamic spectra as structures 
exhibiting a drift in the time--frequency domain. The drift rate depends on their speed and path, 
resulting in a variety of bursts.   Type~III bursts trace the path of supra--thermal electrons 
guided by magnetic structures. They appear, on dynamic spectra, as short (lasting from a fraction of a 
second at dm-waves to a few tens of minutes at km-waves) structures with fast negative 
drift, \citep[$\frac{1}{f} \frac{df}{dt} \approx 0.5  \rm \; sec^{-1}$;~see for example~][]{GuedelBenz88}. 
This corresponds to  anti-sunward propagation of the electrons through regions of decreasing 
ambient density at speeds $\approx c/3$   \citep[e.g.,][]{Suz:Dul-85}. Sunward travelling beams produce 
reverse drift bursts (RS bursts), and beams propagating in closed loops emit type U or J bursts comprising 
a succession of an initial drift towards lower frequencies and a more or less pronounced RS burst. 

Type~II bursts are more slowly drifting bursts
\citep[$\frac{1}{f} \frac{df}{dt} \approx 0.001-0.01 \rm \; sec^{-1}$;~see, for example, Table A.1 in~][]{Caroubalos04}  
excited by electrons accelerated at travelling shocks 
and emitting in their upstream region. Finally broadband dm-m 
wave continuum emission that may last over several minutes or even hours (type IV burst)  is 
ascribed to electrons confined in closed coronal magnetic structures. The reader is referred to the 
reviews in \cite{McLean85}, \cite{Bas:al-98}, \cite{Nindos08} and \cite{Pick08} for more detailed 
accounts of the radio emission by non thermal electrons in the corona.
%-------------------------------------------------------------------------------------
\begin{table}[t]
\centering
\caption{Overview of the 17 January 2005 Event and associated activity. }
\label{T}
\begin{tabular}{{lllll}} %
\hline\hline
\textbf{Event}		&\textbf{Time}  	&\textbf{Characteristics} & \textbf{Remarks}	 \\
					&\textbf{UT}    	&      		 			&			 \\
\hline
SXR Start   			& 	06:59   		&       				& AR10720 (N15$^\circ$ W25$^\circ$) \\
Type IV     			& 	08:40   		& 3.0-630 MHz				& AR10720  	  	 \\
CME$_1$     			&   	09:00			&               			& lift-off             	 \\
\hline
\textbf{SXR Stage 1} 		&	09:05   		&       				&  			 \\
{ First cm  }			&	09:05   		&       				& RSTN 15400 MHz  \\
{  burst start}			&				&					&			\\
Type III$_1$			& 09:07-09:28  			& 0.2-630 MHz  	    			& AR10720  	  	 \\
Type II$_1$ 			&	09:11   		& 0.2-5   MHz  				& AR10720  	  	 \\
\Ha~Start 			&	09:13   		& 3B     	    			& KANZ, AR10720  	  	 \\
CME$_1$     			&   	09:30   		& 2094 km sec$^{-1}$   			& On C2			 \\
HXR Start   			&   09:35:36			&					& RHESSI Number 5011710	 \\
CME$_2$     			&   	09:38   		&               			& lift-off             	 \\
\hline
\textbf{SXR Stage 2} 		&	09:42			&       				& End SXR Stage 1  	 \\
{ Second cm }			&	09:43   		&       				& RSTN 15400 MHz  \\
{  burst start}			&				&					&			\\
Type III$_2$			& 09:43-09:59  			& 0.2-630    	    			& AR10720 \\
HXR peak    			&   09:49:42			& 7865 counts sec$^{-1}$		&             		 \\
Type II$_2$ 			&	09:48   		& 0.2-8 MHz  	    			& AR10720 \\
SXR peak    			&	09:52   		& X3.8   				& End SXR Stage 2  	 \\
\hline
CME$_2$     			&   	09:54   		& 2547   km sec$^{-1}$  		& On C2			 \\
First rise			&	10:00			& 38-315 keV				& ACE/EPAM		\\
Electron flux			&				&					&			\\
SXR End     			&	10:07   		&       				& AR720  	  	 \\
HXR End     			&   10:38:52			& 53152112 total counts			& RHESSI		 \\
Second rise			&  12:00 			& 38-315 keV				& ACE/EPAM		\\
Electron flux			&				&					&			\\
Type III$_3$			&	11:37			& 0.5 MHz    	    			& CME$_1$, CME$_2$ merge at 37 \RSUN \\
				&				&      		 			& type II$_2$ overtakes type II$_1$ \\
\Ha~End	    			&	11:57   		&	        			& KANZ  		 \\
Type IV End 			&	15:24   		& 3.0-630 MHz	    			& AR10720 \\
\hline\hline
\end{tabular}
\end{table}
%-----------------------------------------------------------------------------------------------
\begin{figure}[h!]\centering
\includegraphics[scale=0.70]{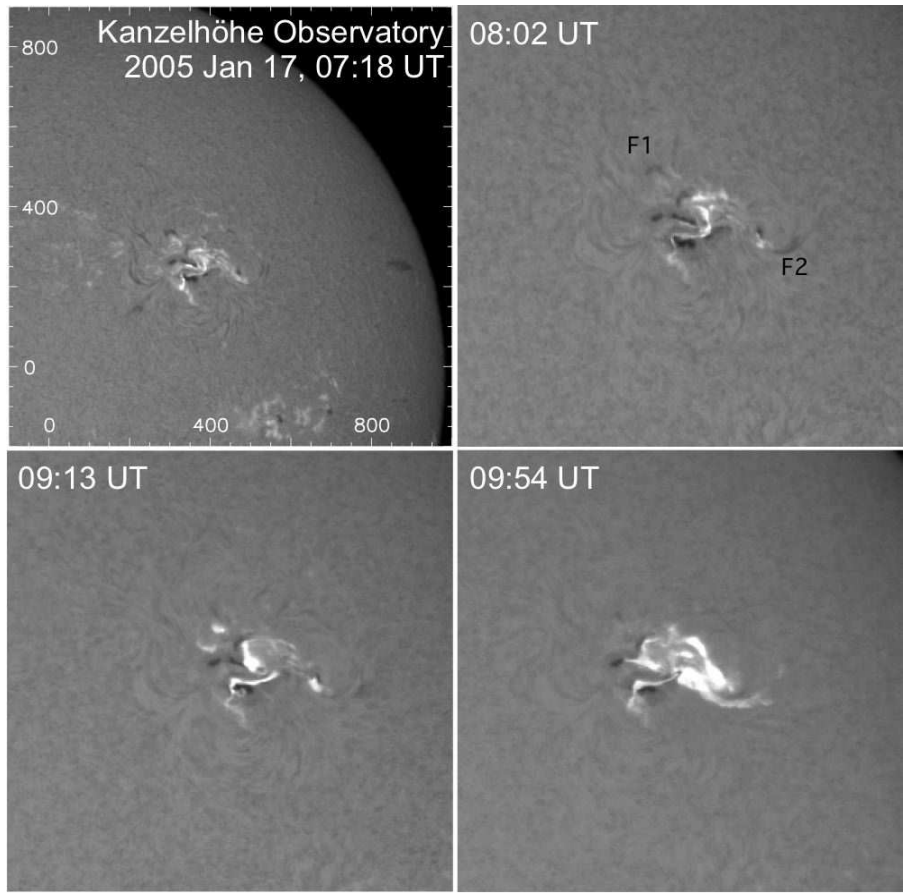}
\caption{Snapshots of active region NOAA 10720 on 17 January 2005 in \Ha  ~line centre (top left) and in the wing, 
observed at Kanzelh\"ohe Observatory (courtesy M.~Temmer). Solar north is at the top, west on the right.
The two snapshots at the top show the active region before the flare under discussion, the two bottom images 
show two instants during the stages 1 and 2, respectively. 
These stages were associated with the disappearance of the filaments labelled `F1' and `F2'.}
\label{Fig_KANZ}
\end{figure}
%-----------------------------------------------------------------------------------------------
\begin{figure}[h!]\centering
\includegraphics[scale=0.95]{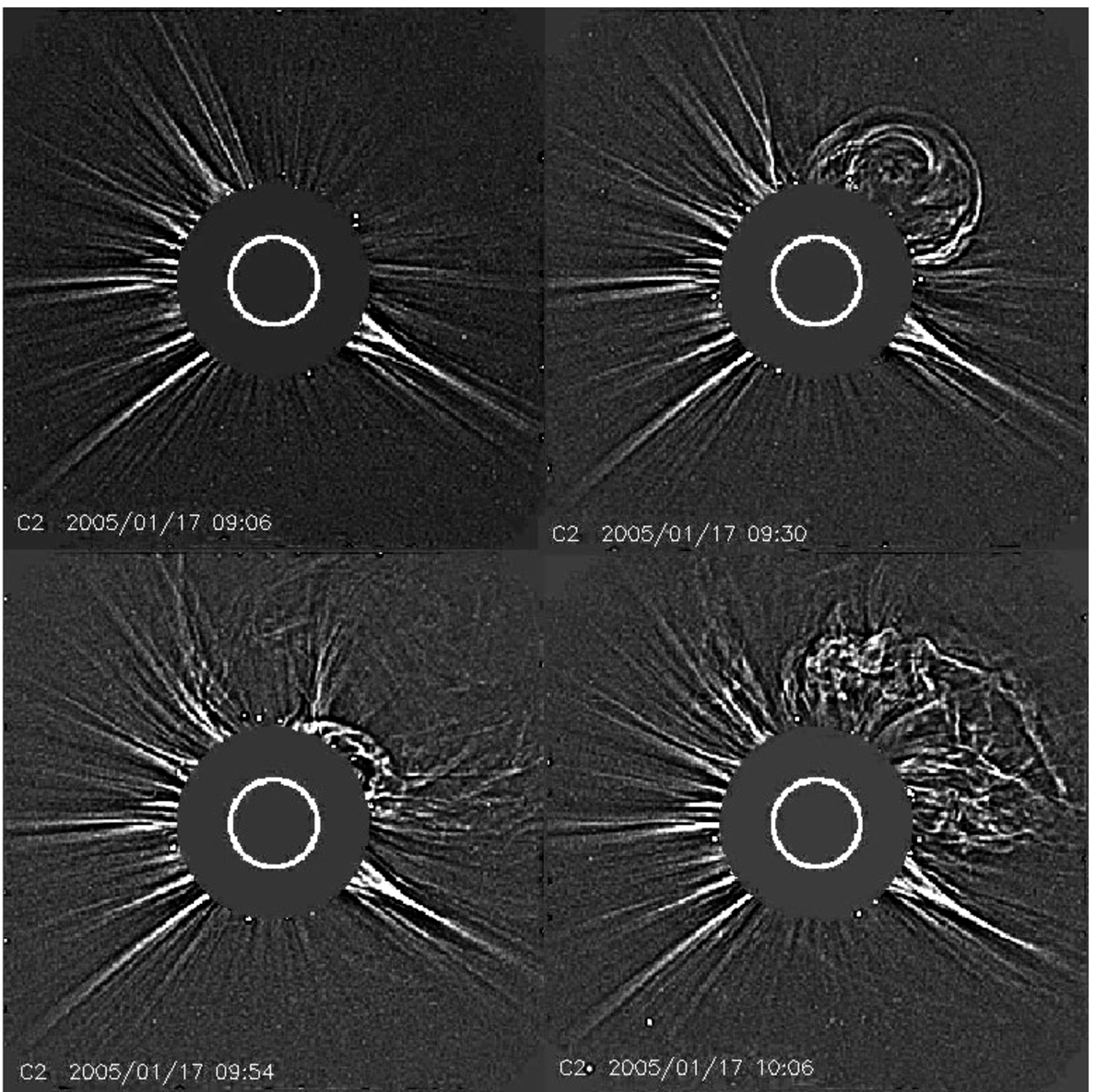}
\caption{The two LASCO CMEs in close succession; the images have been subjected to high-pass filtering. 
Top: Two frames of the 09:30:05 Halo CME with 
back-extrapolated lift off at 09:00:47 UT and plane-of-the-sky speed 2094 km sec$^{-1}$. Bottom: Two frames of the
09:54:05 Halo CME with back-extrapolated lift off at 09:38:25 UT and plane-of-the-sky speed 2547 km sec$^{-1}$.
Solar north is at the top, west on the right.}
\label{CMES}
\end{figure}
%-----------------------------------------------------------------------------------------------
\section{Observations and Data Analysis} \label{Event}
The 17 January 2005 event consisted of a complex flare, two very fast coronal mass ejections (CMEs), and intense and 
complex soft X-ray (SXR) and radio emission. In all radiative signatures two successive stages can be 
distinguished.  
The CMEs  were launched successively from neighbouring regions of the corona and interacted in interplanetary space.  
The sequence of the observed energetic phenomena is summarized in Table \ref{T} and described,
in detail, in the following subsections.

\subsection{Optical observations: flares and CMEs}

Figure~\ref{Fig_KANZ} displays snapshots in the \Ha ~line obtained from the Kanzelh\"ohe solar observatory 
(courtesy M.~Temmer; see also \citet{Temmer2007}, their figure 2, for details on the evolution of the \Ha ~flare
ribbons).  The only major active region on the disk is NOAA~10720 in the north-western 
quadrant (N15$^\circ$ W25$^\circ$). It displayed nearly uninterrupted activity  since the early hours 
of 17 January 2005. The most conspicuous event was a 3B~\Ha ~flare reported by Kanzelh\"ohe 09:16-11:57 UT. 
This flare proceeded successively  in two { different parts of AR 10720},
as shown in the two snapshots of the bottom panel. The first part of the event, referred to as ``stage~1"
(illustrative snapshot at 09:13~UT), is seen in the eastern part
of the active region,~close to the sunspots. It is associated with the 
 temporary disappearance or eruption of the filament labelled `F1' in the upper right panel. 
%------------------------------------------------------------------------------------------------------------
Two major flare ribbons are distinguished in the snapshot at 09:13~UT: {\bl
a narrow band essentially in the east-west direction and a broader north-southward oriented region. The significant offset of the two ribbons with respect to the neutral line shows the eruption of a strongly sheared magnetic field. } 
%------------------------------------------------------------------------------------------------------------
After about 09:35 UT the brightest emission is seen in the western part of the 
active region (``stage~2"; see snapshot at 09:54 UT), 
together with the eruption of another filament  `F2'
(or of a different part of the filament whose northern section erupted before). 
The brightening 
consisted of two essentially parallel flare ribbons, which were connected by post flare loops in later 
snapshots (not shown here).  These two stages of the event were also found in the 
soft X-ray (SXR) and radio emissions, as will be discussed below.

Two CMEs were observed in close succession. A sequence of difference images from the
Large Angle and Spectrometric Coronagraph (LASCO) aboard the SOHO spacecraft \citep{Brueckner95} 
is displayed in Figure~\ref{CMES}: the first CME (henceforth CME$_1$) is seen in the image at 09:30~UT  in the
north-western quadrant. While it travelled through the corona, the second, broader CME (CME$_2$) appeared underneath 
(image at 09:54~UT). The most conspicuous features of both CMEs are seen above the north-western limb, but both were labelled halo CMEs in the LASCO CME 
catalog\footnote{\url{http://cdaw.gsfc.nasa.gov/CME_list/}} \citep{Yashiro}.
%-------------------------------------------------------------------------------------
Speeds of, respectively, 2094 and 2547~km~s$^{-1}$ were derived from linear fits  to the trajectories of their fronts 
{\bl published in the CME catalogue.}
{\bl Formally the CME fronts described by the fits intersected near 12:32~UT at a heliocentric distance of about 38~\RSUN. }
%-------------------------------------------------------------------------------------
The statistical error of the speeds of the CME fronts and their liftoff times, derived from the abovementioned linear 
least-squares fit to the measured heliocentric distances, leads to an uncertainty of $\pm$3~h in the time of intersection. 
This uncertainty  stems from the fact that the two height-time trajectories are nearly parallel. 
{\bl We will show in Sect.~\ref{CME2CME} that CME interaction actually occurred well before the formal time of intersection.}
Of course a single instant of interaction between two complex CMEs is fictitious anyway.
%-------------------------------------------------------------------------------------
% a landscape figure which scales and rotates the file untitled01B.eps
\begin{sidewaysfigure}
\centering
\resizebox{\textwidth}{!}{\includegraphics{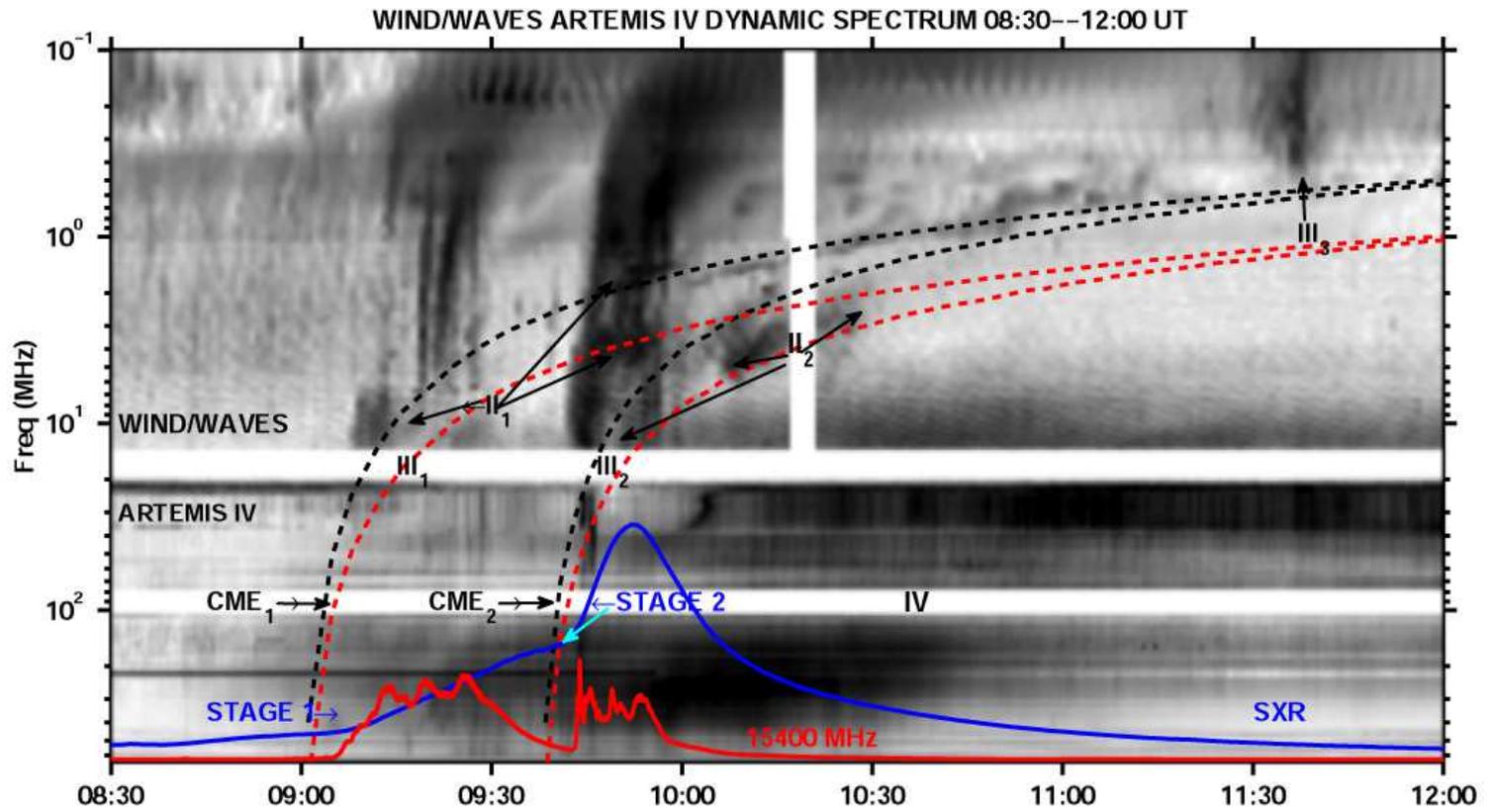}}
\caption{ARTEMIS-IV/{\it Wind}/WAVES dynamic spectrum (inverse grey scale). Overlays:
The profiles of GOES SXR flux (dark blue) and RSTN flux density at 15400 MHz (red) 
and the frequency-time plots  derived from the linear fits to the front trajectories of CME$_1$ and CME$_2$ and 
an empirical density model for fundamental (black dashed curve) and harmonic (red dashed curve) plasma emission.
The type IV continuum, the type II (II$_1$ and II$_2$) and type III GG bursts (III$_1$ and III$_2$), the 
stages 1 and 2 of the SXR flux rise, and the type III burst (III$_3$)
around the convergence of the fronts of CME$_1$ and CME$_2$ are annotated on the plot.}\label{F3}
\end{sidewaysfigure}
%-----------------------------------------------------------------------------------------------

\subsection{Soft X-ray and radio emission}\label{SXR_Radio}%ALEX% Added a LABEL Here
An overview of the complex radio event is given in Figure~\ref{F3}. There we present the  dynamic flux density spectrum of the radio bursts 
in the 650~MHz-20~kHz range (heliocentric distance $\approx$ 1.1 \RSUN to 1 AU)
using combined recordings of the {\it Appareil de Routine pour le Traitement 
et l'Enregistrement Magn\'etique de l'Information Spectrale} (ARTEMIS-IV)  solar radio-spectrograph
\citep{Caroubalos01,Kontogeorg06} and the {\it Wind}/WAVES experiment \citep{Bougeret95}.
Several  other time histories are superposed on the dynamic spectrum:
\begin{itemize}
  \item { 
{\bl Dashed lines display the approximate frequency-time trajectories of the two  CME fronts, using the density model of  \citet{Vrsnak04},
which describes well the coronal density behavior in the large range of distances from low corona to interplanetary space}: %KLK
%---------------------------------------------------------------------------------------------------
 \begin{displaymath}
 \frac{n}{10^8 \; \rm cm^{-3}}=15.45 \left( \frac{R_\odot}{R} \right)^{16}  
                                                    + 3.165 \left( \frac{R_\odot}{R} \right)^6  
                                                    + 1.0 \left( \frac{R_\odot}{R} \right)^4  
                                                    + 0.0033 \left( \frac{R_\odot}{R} \right) ^2  
 \end{displaymath}

{\bl The linear fits to the height-time trajectories of the CME fronts in the LASCO images were converted to frequency-time 
tracks of fundamental (black line) and harmonic (red line) plasma emission.} }
%---------------------------------------------------------------------------------------------------
  \item {The  solid blue curve displays the SXR time history (0.1-0.8 nm), 
	using GOES on line data (\url{http//www.sel.noaa.gov/ftpmenu/indices}),
	describing thermal emission from the flare-heated plasma.}
  \item {The red curve is the microwave time history at 15.4~GHz, produced by non thermal electrons (energies $>$100~keV) 
	in magnetic fields of a few hundred~G; these were obtained from the San Vito Solar Observatory of the 
	Radio Solar Telescope Network (RSTN) \citep{Guidice81}\footnote{ \url{http://www.ngdc.noaa.gov/nndc/struts/form?t=102827&s=6&d=8,40,9}}.}
\end{itemize}
The two stages of the flare identified in the \Ha ~observations in Figure~\ref{Fig_KANZ}  correspond to two 
distinct events of energy release seen in the SXR and microwave time profiles (Figure~\ref{F3}). The SXR time profile had an 
initial smooth increase between 06:59~UT and 09:05~UT.  Subsequently the SXR flux rose slightly faster until 09:45~UT (stage 1), 
and even faster (stage 2) until the X3.8 peak at 09:52~UT.  The gradual rise in stage 1 and the faster rise in stage 2 were each accompanied by strong microwave bursts. 
The second burst was also observed in hard X-rays by 
RHESSI\footnote{ \url{http://hesperia.gsfc.nasa.gov/getfiles/rhessi_data_search.html}}~\citep{Lin02}\footnote{
{During the stage 1 of the SXR flare RHESSI was in the Earth's shadow, so there is no HXR burst associated with the 
first microwave burst.
}}.
%
%-------------------------------------------------------------------------------------
\subsection{Radio emission: decametre-to-kilometre waves}\label{DM}
%-------------------------------------------------------------------------------------
The dominant features in the dynamic spectrum observed by {\it Wind}/WAVES
 at frequencies  below 2~MHz are two groups of type~III bursts, labelled III$_1$ and III$_2$. They occurred in 
association with the SXR and microwave emissions of stages 1 and 2, respectively, and with the two different parts of 
the {\Ha}  flare.  The two type~III groups occurred near the extrapolated liftoff times of the two CMEs.  
Radio images taken by the Nan\c cay Radioheliograph \citep[NRH;][]{Kerdraon97} show that the sources are 
located in the north-western quadrant near  the flaring active region\footnote{We have used 164--432 MHz contours 
overlaid on EIT 19.5 cm images}. Hence both flare episodes were efficient accelerators of electrons that escaped to 
the interplanetary space along open magnetic field lines rooted at or near the flare site. The second type~III 
group (Type III$_2$) was followed by a more slowly drifting narrow-band burst  (type II, labelled II$_2$) produced 
by a coronal shock wave. Upon closer inspection the spectrum suggests that similar drifting features can also be 
associated with the first flare {episode}, although the association is less evident. We label these bursts II$_1$ in Figure~\ref{F3}. 
{Since the two CMEs are extremely fast, they are expected to drive shock waves in the corona.  {\bl The observed type II emission can be 
compared with the dashed curves in Figure \ref{F3}, which track fundamental (black) and harmonic (red) emission expected from the 
trajectory of the CME front and the coronal density model.} It is clear that this density model is only indicative, especially 
in the perturbed corona through which travels the second CME } (see discussion in subsection \ref{M}).   
We therefore associate type II$_1$ and II$_2$ to the bow shocks of the two CMEs, although other 
interpretations, like shocks on the flanks or shocks from a driver related to the flare, are not excluded.

%-------------------------------------------------------------------------------------
\begin{figure}[h!]\centering
\centering 
\resizebox{\textwidth}{!}{\includegraphics{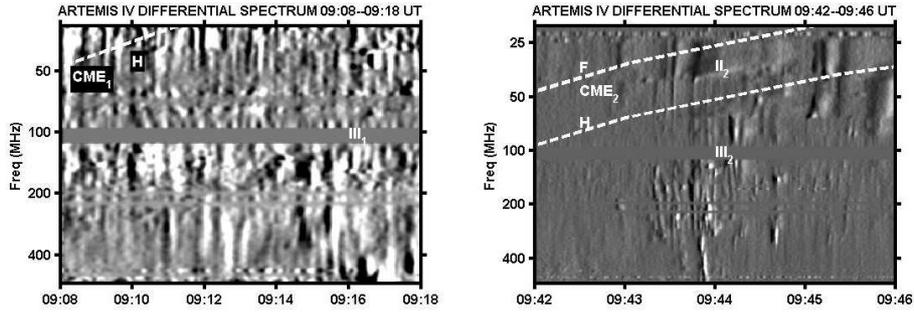}}
\caption{ARTEMIS IV differential spectra in the 20-550 MHz range. 
Top: Type III$_1$ in the 09:08-{09:18} UT interval. Some bursts of the type III family ({\bl Reverse slope (RS),
J--type, U--type}) have been annotated. ~Bottom: Type III$_2$ in the 09:42-09:46 UT interval. The frequency-time trajectories 
of CME$_1$ and CME$_2$ (white) are overlaid, assuming fundamental (F) and harmonic {\bl plasma emission}.}
\label{F4}
\end{figure}
%-------------------------------------------------------------------------------------
\begin{figure}[h!]\centering
\centering 
\resizebox{\textwidth}{!}{\includegraphics{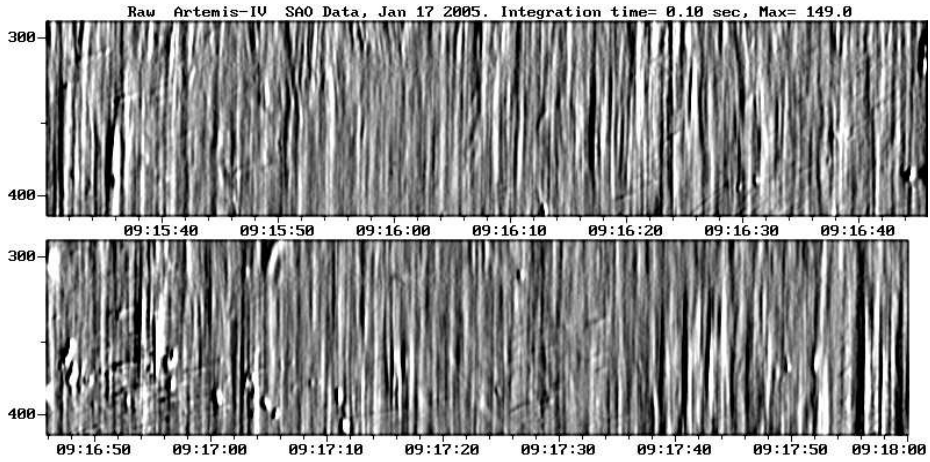}}
\caption{{ARTEMIS IV high resolution (SAO) differential spectra in the 290-415 MHz range and the 09:15:30-09:18:00 UT interval. }}
\label{FS1}
\end{figure}
%-------------------------------------------------------------------------------------
\begin{figure}[h!]\centering
\centering 
\resizebox{\textwidth}{!}{\includegraphics{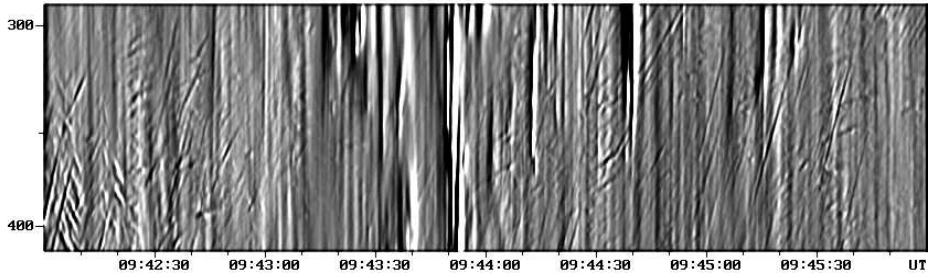}}
\caption{{ARTEMIS IV high resolution (SAO) differential spectra in the 290-415 MHz range and the 09:42-09:46 UT interval. }}
\label{FS2}
\end{figure}
%-------------------------------------------------------------------------------------
\subsection{Radio emission: metre waves}\label{M}
%-------------------------------------------------------------------------------------
{The dm-m wave emission consisted of a type~IV continuum, the metre wave counterparts of the dekametre-hectometre (DH)  
type~III groups and of the {\bl type II bursts}. 

The  type~IV continuum started near 08:40~UT during the initial smooth increase of the SXR flux 
before  stage~1. It was first visible as a grey background in the dynamic spectrum, and became progressively more intense. 
It dominated the metre wave spectrum during and after Type III$_2$,  and  gradually penetrated to lower frequencies, down to 5~MHz. 
%-------------------------------------------------------------------------------------
Images in the   EIT 195 \AA~channel \citep{Delaboudiniere95} and in the 164-432 MHz range 
taken by the NRH indicate that the thermal 
{\bl (soft X-rays) }  and non thermal {\bl (radio) }
emissions all originated near NOAA AR~10720. In the time interval from the start of the type IV 
burst to the start of stage 1 a wealth of fine structures was recorded \citep[see][]{Bouratzis09}.
From the high-resolution observations in the 200-500 MHz range 
{(see figures \ref{FS1}, \ref{FS2} for example) }
it appears that most bursts are broadband pulsations. Other fine structures of type~IV emission 
such as spikes, fiber bursts and zebra pattern appear 
occasionally \citep[see][for a description of fine structure of type~IV 
emission]{Kuijpers80}. }

During type III$_1$ the spectral character of the radio emission was clearly different at 
frequencies below and above the inferred frequency-time track of the CME$_1$
{\bl (see Fig.~\ref{F3}). }
On the low-frequency side of the track strong type III bursts 
were prominent after about 09:22~UT. They were preceded by a less regular 
emission, which \citet{Reiner08} label ``complex type~III bursts" because of 
{\bl its }% 
 varying flux density across the spectrum. 

The metre wave counterpart on the high-frequency side of the estimated CME track  
{\bl consisted of a succession of spectral fine structures on the time scale of seconds, 
with different spectral characteristics superposed on the type~IV continuum, followed after 09:11~UT
by the high-frequency extension of the dekametre-hectometre (DH)  type~III group III$_1$.  
A more  detailed view of the difference spectrum is given in the top
panel of Figure~\ref{F4},  while high resolution images of the fine structures are in figure \ref{FS1}. 
Among these fine structures were broadband pulsations, bursts with ordinary and reverse drift, and fiber 
bursts due to whistlers travelling upwards in the corona (see Figure \ref{FS1}, {\it e.g.}, 09:16:20-09:16:45~UT). 
The variety of these bursts shows the acceleration and partial trapping of  electron populations in the 
corona well behind the front of the CME. Indeed, few of }
the well-identified bursts above 100~MHz seem to continue into the 30-70~MHz range.  
It was only near the end of type III$_1$ ($\approx$ 08:18 UT) that metre wave type~III 
bursts appeared as systematic high-frequency extensions of the type~III bursts observed below 2~MHz. 

Type III$_2$ started at 09:43 UT, together with the second microwave burst, near the 
back-extrapolated lift-off of CME$_2$ (09:38 UT) and  the onset of stage 2  of the SXR burst. 
A close look at the dynamic spectrum ({bottom} panel of Figure~\ref{F4}) reveals 
{\bl
negative overall drifts below 100~MHz, while burst groups with positive overall drift prevailed 
above 130~MHz. The high-resolution spectrogram in the 300-400 MHz range (Fig.~\ref{FS2}) shows a 
wealth of individual bursts with different drift rates and zebra pattern. These bursts show again, 
like in stage~1 of the event, that the high-frequency bursts are produced by an accelerator below 
the CME front, while low-frequency bursts show the start of the prominent DH type~III bursts.}
%
%-------------------------------------------------------------------------------------
\subsection{Radio emission and CME interaction}\label{CME2CME}
%-------------------------------------------------------------------------------------
\begin{figure}[h!]
\centering
\includegraphics[width=0.55 \textwidth, height=0.25 \textheight]{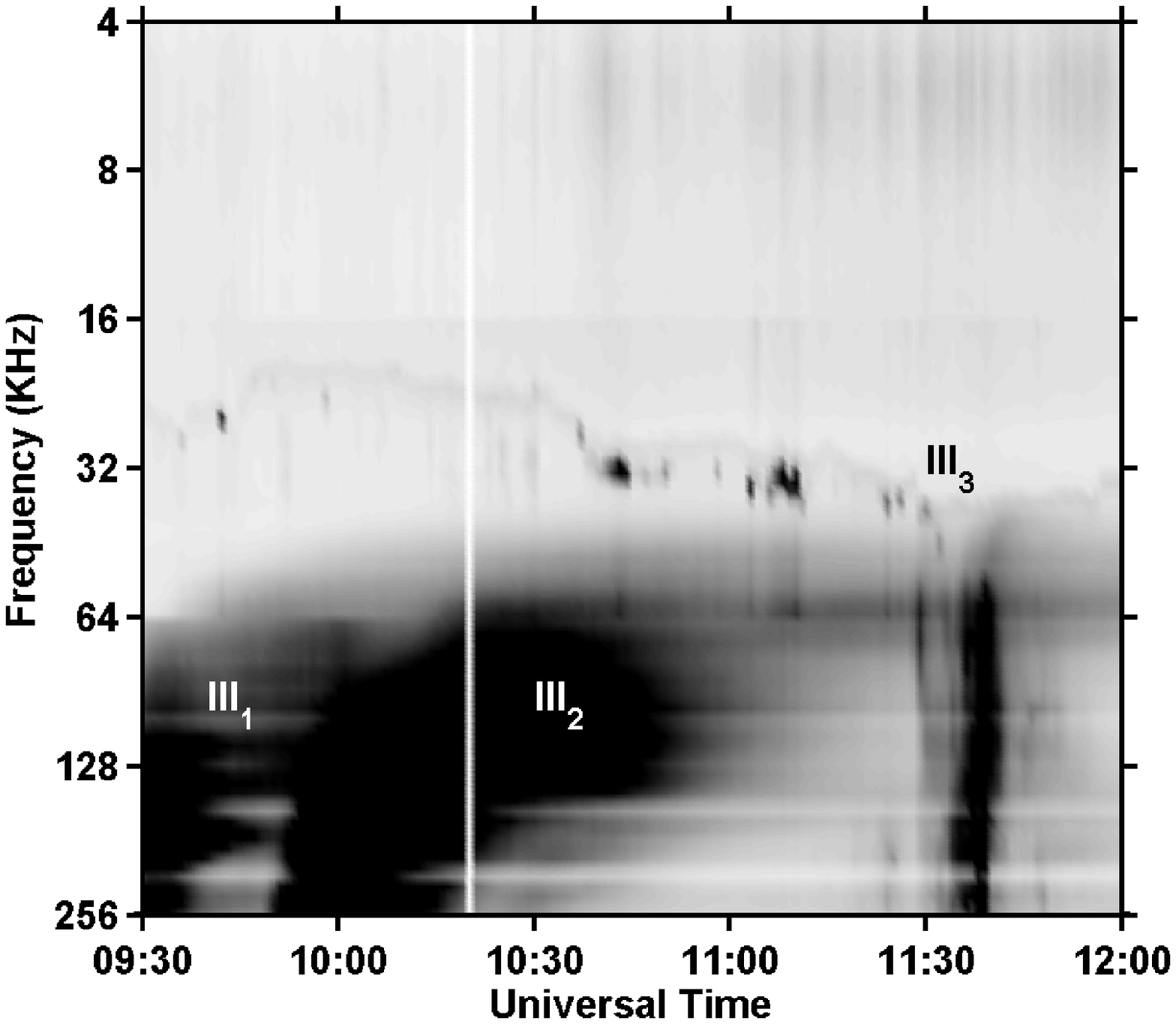}
\includegraphics[width=0.40 \textwidth, height=0.25 \textheight]{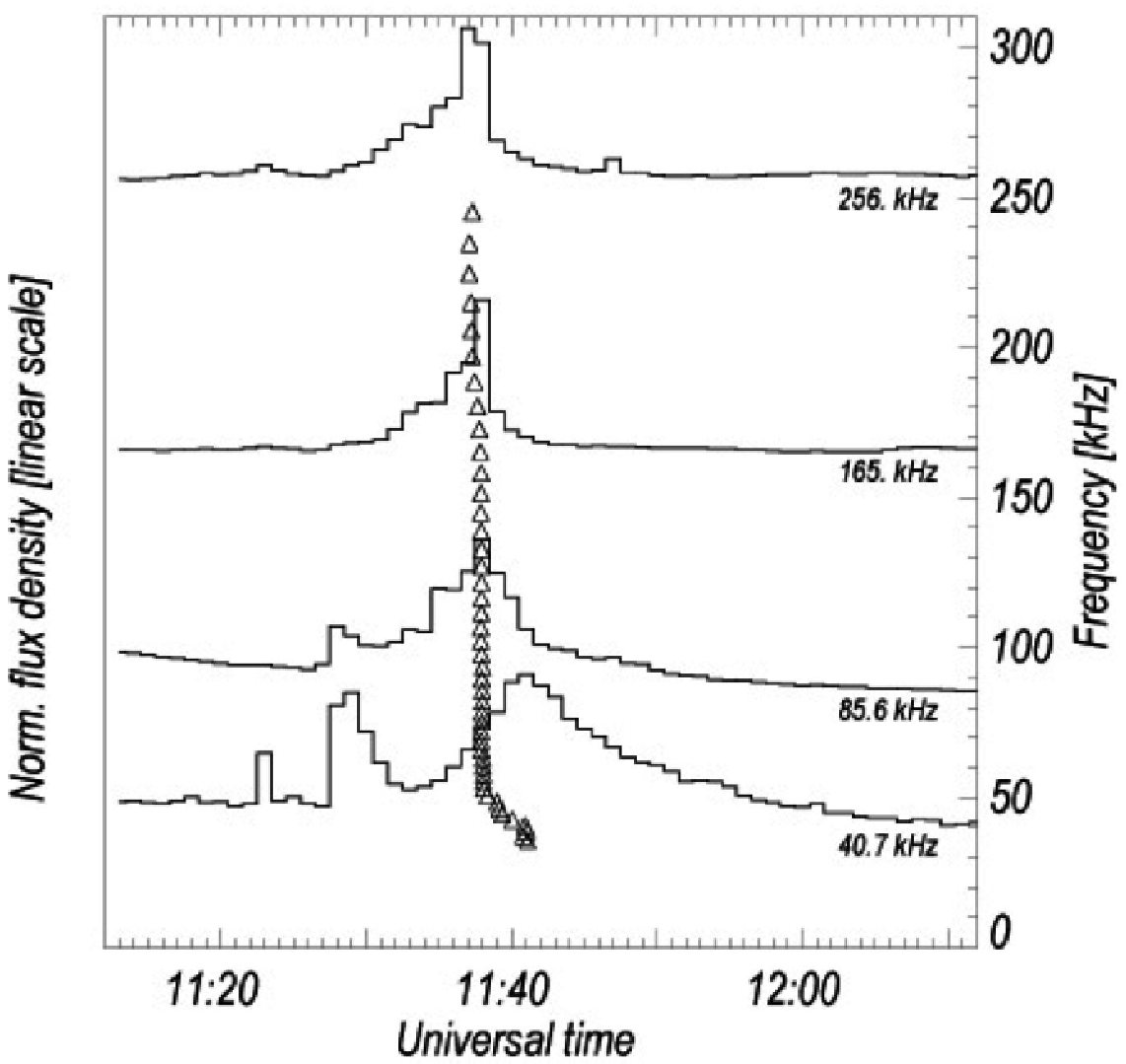}
\caption{
Kilometric radio bursts associated with the flares and with CME interaction.
Left: WAVES/TNR dynamic spectrum (inverse grayscale) in the 4-256 kHz frequency range 
from 09:30~to 12:00 UT showing the burst groups  type III$_1$, III$_2$ and III$_3$.
Right:  Time histories during type III$_3$ (each normalised to its maximum) at selected frequencies. 
The open triangles show the time of maximum of the burst at each frequency in the range 35-256 kHz.}
\label{Fig_TNR_20050117}
\end{figure}
%-------------------------------------------------------------------------------------
\begin{figure}[h!]
\resizebox{\textwidth}{!}{\includegraphics{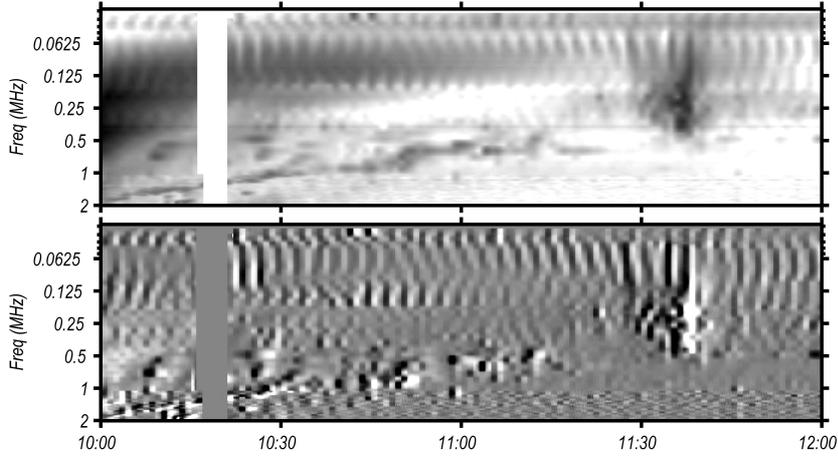}}
\caption{{\it Wind}/WAVES  Spectra in the 2.0-0.03 MHz range with details of 
the merged type II bursts (10:00-11:20 UT) and the low-frequency type III (III$_3$);
about 11:30-11:40 UT). Top: Intensity, Bottom: Differential.}
\label{III3}
\end{figure}
%-------------------------------------------------------------------------------------
Well after the decay of the SXR and microwave emission a third group of bursts 
(III$_3$ in Figure~\ref{F3}) is identified (near 11:30 UT), with unusually low 
starting frequency (0.6 MHz), pointing to an acceleration of the emitting electrons at unusually 
great height.  
A more detailed view of the low-frequency radio spectrum of this burst group and the preceding groups III$_1$ 
and III$_2$ is given by the dynamic spectrum as observed by the {\it Thermal Noise Receiver} (TNR) of {\it Wind}/WAVES 
in the left panel of Figure~\ref{Fig_TNR_20050117}. The narrow-band short bursts near 32~kHz are Langmuir wave packets. 
Together with the fainter continuous band on which they are superposed they indicate how the electron plasma frequency 
evolves at the {\it Wind} spacecraft. At the time of III$_3$ it is about 35~kHz. Using a standard interplanetary density model, 
where the electron plasma frequency decreases as the inverse of the heliocentric distance ~$R$, the starting frequency 
of III$_3$ implies $R=12$~{\RSUN} for fundamental plasma emission, and $R=25$~{\RSUN} for the harmonic. From the LASCO 
observations and the uncertainties resulting from the straight-line fits to the CME front trajectories, the heliocentric 
distances of the CME fronts at 11:30~UT are, respectively, $(26.9 \pm 4.8)$~{\RSUN} and $(24.5 \pm 5.5)$~\RSUN.  The burst 
group III$_3$ is hence consistent with harmonic emission from the vicinity of the CME fronts.  
This points to a close relationship of this episode of electron acceleration with the interaction of the two CMEs. 

Comparison of the three groups of type~III bursts in the TNR spectrum of Figure~\ref{Fig_TNR_20050117} shows 
that type III$_3$ is much shorter than the previous type~III bursts. It has intrinsic structure 
that indicates a group of bursts. The low-frequency cutoff is 
near the plasma frequency at the spacecraft at that time.
More details are seen in the selected time profiles in the right panel, plotted
together with the peak times of the burst at each frequency in the 35-256 kHz range (open triangles).  
The time profiles show that the peak times are not distinguishable over a large part of the frequency 
spectrum with 1-min integrated data, but that the centre of gravity of the brightest feature
shifts to later times at the lower frequencies.  We determined the maximum of the burst at each of the TNR frequencies where 
it is well defined, using  a parabolic interpolation between the observed maximum and its two neighbours. 
It is this interpolated time which  is plotted by an open triangle. The peak time spectrum resembles a type III burst 
especially at the lower frequencies. The peak time delay  
is merely 1~min between 250 and 50~kHz, but becomes  clear at frequencies below 50~kHz. 
For comparison, the peak time delay between 50 and 250~kHz is 42 min
during the previous burst III$_2$. 
The frequency drift rate is hence faster than 3~kHz/s during III$_3$, as compared to 0.08~kHz/s ~during III$_2$. 
Because of the morphological similarity in the dynamic spectrum, and despite the different drift rates, we 
assume in the following that the type III$_3$ bursts are indeed produced by electron beams travelling in 
the anti-sunward direction from the acceleration region.
Since the emission extends rapidly to the plasma frequency at the spacecraft, we conclude that the electron beams do not travel within the CMEs, 
but escape rapidly from the acceleration region in the vicinity of the CME fronts to 1~AU. This means that they must travel along pre-existing open 
solar wind field lines.

{\bl
To the extent that drift rates reflect the speed of the exciter, the fast frequency drift of the type III$_3$ bursts 
implies that the exciter speed is higher than during the preceding groups III$_1$ and III$_2$. }

The total and differential radio spectrum observed by {\it Wind}/WAVES 
are shown in Figure~\ref{III3}. The spectrum shows a chain of narrow-band emissions with negative frequency drift, indicating the type~II bursts, 
followed by the high-frequency part of type III$_3$ between 11:28 and 11:40 UT. The spectrum in Figure~\ref{F3} leaves it open if  this is the 
continuation of the first type~II burst ({II$_1$}), presumably associated with CME$_1$, 
or whether it contains contributions from both CMEs. 
The starting frequency of the type~III bursts is similar to the type~II frequency when extrapolated to the time
of the type~III bursts. This is consistent with the type III electron beams radiating
%implies  that  the type~III emitting electron beams radiate 
in the upstream region, like the shock-acclerated electrons emitting the type~II burst.
One may go one step further and consider this coincidence as a hint that the  electron beams are accelerated at the shock,  as argued in cases where 
type~III bursts clearly emanate from type~II lanes \citep[see~][]{Bougeret98,Mann}. 
 We will come back to this problem in the Discussion. 
%-------------------------------------------------------------------------------------
\subsection{Solar wind and energetic electrons near 1 AU}  
\begin{figure}
\centering
\includegraphics[width=0.9  \textwidth, height=0.25 \textheight]{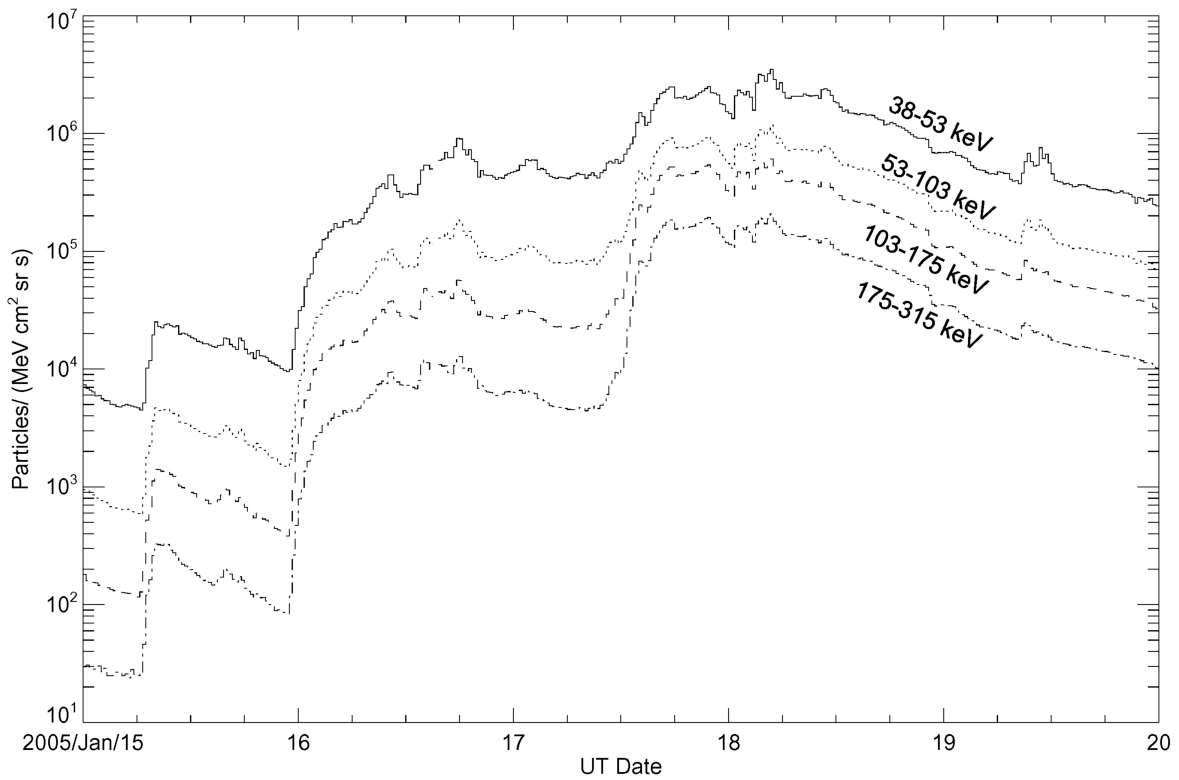}
\includegraphics[width=0.9  \textwidth, height=0.30 \textheight]{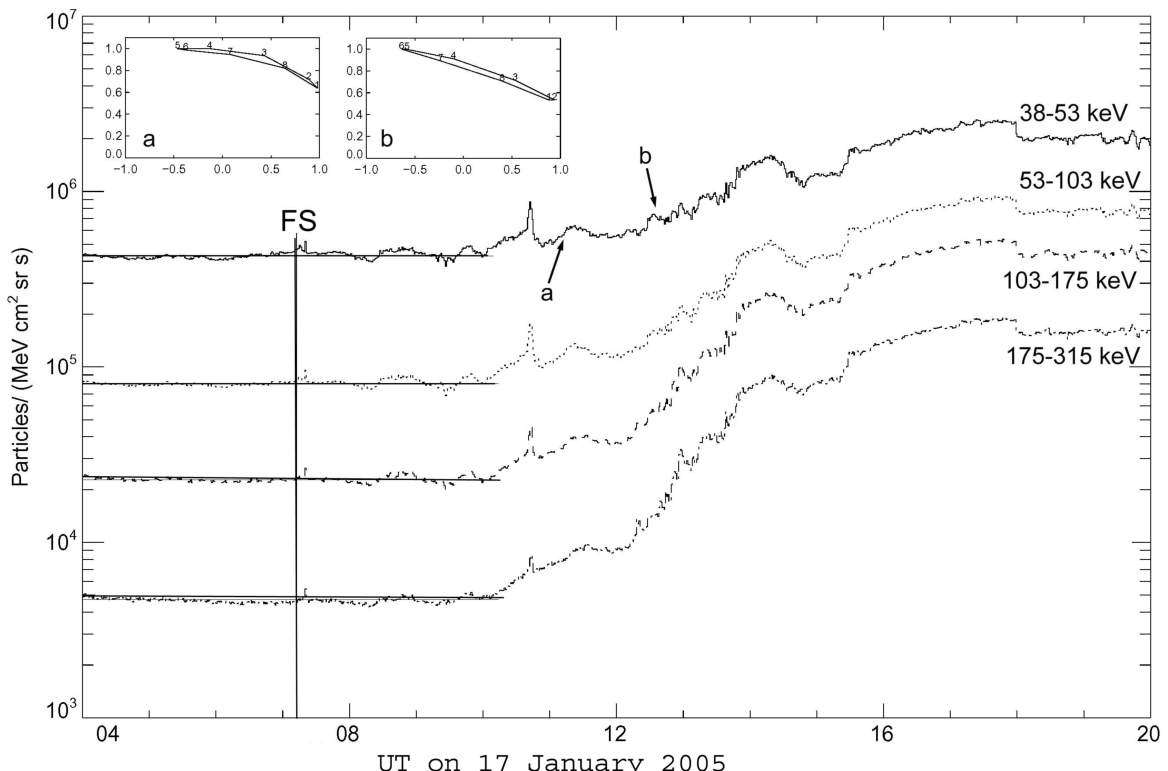}
\caption{
ACE/EPAM energetic electron enhancements in response to the 17 January 2005 solar 
activity. Top: Twenty-minute averaged differential intensities for the interval January 15-20, 2005. 
Labels of the abscissa are days. Bottom: 1-min averaged intensities (38-315 keV) for the interval 
04:00-20:00 UT on 17 January 2005 of four magnetically deflected electron channels (38-315 keV) from 
the B detector head of the EPAM/CA60.  Labels of the abscissa give the day (17 January) and the hour.
The horizontal lines denote pre-event ambient intensity. 
The solid vertical line marks the arrival of an interplanetary forward shock (FS) at 07:12 
UT at ACE. Inserts {\bf a} and {\bf b} present pitch angle distribution  snapshots
at the onset of the event at about 11:00 UT and at about 12:10 UT;
normalized differential electron intensity is plotted versus the cosine of the pitch angle.}
\label{O1}
\end{figure}

%-------------------------------------------------------------------------------------
The energetic particle data were obtained from the {\it Advanced Composition Explorer} (ACE) spacecraft. We use high-resolution 
intensities of magnetically deflected electrons (DE) in the 
energy range 38-315 keV measured by the B detector of the CA60 telescope of the EPAM 
experiment \citep[{\it Electron, Proton and Alpha Monitor};][]{Gold98} on board ACE, and measurements of the angular distributions 
in the energy range 45-312 keV detected by the sunward looking telescope LEFS60. In  figure~\ref{O1} (Top) an overview of the 20-min 
averaged differential intensities of four channels
%magnetically deflected electron channels (38-315 keV) as counted by the B detector head of the CA60 
%telescope of the EPAM experiment 
is presented for the interval 15-20 January 2005. AR~10720 produced 
numerous solar events prior to as well as on 17 January 2005 \citep{Malandraki2007,Papaioannou2010}; in response 
to this solar activity, a sequence of energetic electron intensity enhancements was observed. The electron 
intensities are observed to reach their maximum values during this period following the solar events 
on 17 January 2005. 
%-------------------------------------------------------------------------------------
%-------------------------------------------------------------------------------------
Figure~\ref{O1} (Bottom) shows 1-min averaged deflected electron intensities (38-315 keV) 
for the time interval 04:00-20:00 UT on 17 January 2005. The intensities measured during the time 
interval 04:00-08:00 UT for each electron channel have been averaged to obtain a pre-event 
background (denoted by horizontal lines in Figure~\ref{O1}). We defined the onset time of the 
event at ACE for all energy channels as the time when the intensities get $> 2\sigma$ above the 
background and continue to rise from then on. Using this criterion, we found the first significant 
rise of the electron intensities to occur  at 10:00~UT. 
No velocity dispersion was observed, probably because 
the high pre-event ambient intensities (see top panel Figure~\ref{O1}) 
 mask the onset of the electron event \citep[see ][ for a similar case]{Malandraki2005}. 
The spiky increase observed at about 10:40 UT is probably due to X-ray 
contamination. We found no evidence of a magnetic structure influencing the intensity profiles, which 
indicates the observed time intensity changes are not due to spatial structures crossing over the 
spacecraft, but are most likely dominated by temporal effects. 

Twenty-minute averaged representative snapshots of pitch angle distributions (PADs)
are  shown as inserts in the bottom panel of
Figure~\ref{O1}. Normalized differential electron intensity is plotted versus the cosine of the 
pitch angle. Statistically significant PADs are detected first at about 11:00 UT. The PAD snapshot denoted 
as {\bf{a}} in Figure~\ref{O1} indicates that {immediately} after the onset of the event unidirectional %ALEX% For Emphasis
electron anisotropies are observed. 

Based on the observations available we cannot distinguish whether the electrons were directed sunward or 
antisunward, since the magnetic field (not shown) was directed dominantly transverse to the radial direction 
during this period \citep[see Figure 7 in~][~for a similar case]{Lario04}. However, it is highly likely that 
the observed electrons are streaming away from the Sun in response to the intense solar activity
during this period. Furthermore the type III bursts clearly indicate that electrons stream away from the 
Sun (towards regions of lower density). We cannot be certain that the electron population
measured at ACE/EPAM is the high-energy counterpart of the electron beams
emitting the radio waves, yet the overall timing suggests this.

In the work by \citet{Papaioannou2010} a detailed analysis of the plasma and magnetic field measurements 
at 1 AU by ACE during the period 16-26 January 2005
 was carried out. This includes the period under study in 
the present paper. A forward shock was detected at 07:12 UT on 17 January 2005. We have denoted the arrival time of 
this shock by a vertical solid line in Figure \ref{O1} (Bottom). The analysis has shown that after the passage of this 
shock an unusually extended region exhibiting sheath-like characteristics is observed for~$\approx$~1.5 day 
with highly variable magnetic field magnitude and directions and typical to high proton temperatures 
\citep[see Figure 3 of~][~also Ruth Skoug, ACE/SWEPAM PI team, private communication]{Papaioannou2010}. 
This region is probably related to two CMEs ejected in 
close temporal sequence at the Sun on 15 January \citep[see Figure 2 of~][]{Papaioannou2010}.
Subsequently, at $\approx$ 23:00 UT on 18 January 2005 the arrival of an ICME at Earth is detected, 
ending at about 02:30 UT on 20 January.

The energetic electrons observed at 1 AU analyzed in this work were thus detected in the region with 
disturbed magnetic field characteristics following the shock on 17 January 2005. For the purposes of this 
work, as an approximation, we calculated that 
the nominal Parker spiral for the measured solar wind speed of 620 $\rm km \; s^{-1}$ (ACE/SWEPAM)
at the time of the rise of the electron intensity had a length of about 1.05 AU and was rooted near 
W 37$^\circ$ on the hypothetical solar wind source surface at 2.5~\RSUN. This longitude is not contradictory with an active region 
at W 25$^\circ$, because non-radial coronal field lines can easily establish a connection \citep{Klein08}.  
Supposing that the early rise of the intensities was produced by the faster electrons in an 
energy channel moving along the interplanetary field line with 0 pitch angle, we estimate a travel time of 
about 15~min, which indicates the electrons were released  from about 
09:45 UT at the Sun. This corresponds to a photon arrival time at 09:53~UT. Given that our estimate of the electron rise 
gives only an upper limit, we consider that this electron release is related to type III$_1$ 
and type III$_2$ (Table \ref{T}), but cannot give a more detailed identification.
The electron intensities are subsequently observed to exhibit a significant and more abrupt rise at all energies. 
Extrapolation of this second rise to the pre-event background intensities indicates a start 
at $\approx$ 12:00 UT. The electron PADs (inset {\bf{b}} in Figure~\ref{O1}) indicate stronger 
unidirectional anisotropies are observed in association with this electron enhancement, which provides 
evidence for fresh injection of energetic electrons between the Sun and the spacecraft. 

The outstanding radio emission near this time is the group of fast type~III bursts during the CME 
interaction, type III$_3$. 
{\bl If the electrons are accelerated at a heliocentric distance of about 25 \RSUN, the path travelled to the spacecraft 
along the nominal Parker spiral is 0.92 AU for the solar wind speed measured at the time of type III$_3$ (800 km/s). 
The inferred upper limit of the solar release time is 11:46 UT for 100 keV electrons.}
Photons released at that time at 25 \RSUN (0.12 AU) will reach the Earth about 7 min later. 
Since the high background implies that our estimations of the electron rise times are upper limits, 
we consider that this timing is consistent with the type III$_3$ burst group near 11:37 UT (Table \ref{T}).  
This process is hence accompanied by the acceleration of copious amounts of electrons that escape to 
the vicinity of the Earth.

%---------------------------------------------------------------------------------------
\section{Discussion and Conclusions} \label{SandD}
%--------------------------------------------------------------------------------------
On 17 January 2005 two flare/CME events occurred in close temporal succession in the same active region. 
Both CMEs had very high projected speed, above 2000~km~s$^{-1}$, but the second one was faster than 
the first and eventually overtook it. The CMEs were associated with two successive
filament eruptions and ~SXR enhancements in the same active region. 
Since the filament eruptions occurred at neighbouring places in the parent active region, the 
CMEs probably resulted from the eruption of neighbouring parts of the same overall magnetic
configuration. The soft X-ray characteristics of the two successive events were different: 
a  slow monotonic rise to moderate flux during the first event, and a more impulsive rise to 
the X3.8 level ($3.8 \times 10^4 \; \rm W \; m^{-2}$) during the second. Both events had a conspicuous microwave 
burst, but the first one was stronger than the second, contrary to the soft X-rays. The second burst was also seen 
in hard X-rays by RHESSI, which was in the Earth's shadow during the first burst. 

%---------------------------------------------------------------------------------------
\subsection{Evidence for evolving acceleration regions in the corona during the flares}
%---------------------------------------------------------------------------------------
The decametre-to-hectometre wave spectra in the two stages looked similar, with bright groups of type~III bursts signalling 
the escape of electron beams at heliocentric distances beyond 2~$R_\odot$. But radio emission from  lower heights shows 
distinctive differences that point to an evolving acceleration region, {\it i.e} either an acceleration region  which progresses 
through the corona or a number of acceleration sites activated in succession. 

Shock acceleration was clearly at work during both stages of the flare, as shown by the type~II emission. The strong type~III bursts 
at the low-frequency side of the estimated spectral track of the CME front could also be ascribed to the acceleration at the type~II shock. 
The presence of type~III bursts with negative drift at higher frequencies and the fine structures of the type~IV continuum show, however, 
that at the time when the shock traveled through the high corona, other acceleration regions were active at lower altitude, as is usually the 
case during complex type~III bursts at decametre and longer waves \citep[see][and references therein]{Reiner08}. The type~III  bursts (III$_1$)
might then not start near the CME front, but at higher frequencies, and be interrupted by interactions of the electron beams with the turbulence near 
the front of CME$_1$. This is a frequently quoted interpretation of complex features in type~III bursts, both at kilometric \citep{McD-89} and 
decametric wavelengths \citep[e.g.][]{Rei:Kai-99}.

In the second type III group ~{(III$_2$)} the overall frequency drift of the low-resolution spectrum (Figure \ref{F4}) was positive.
The persistence of the metric type~IV burst, which suggested acceleration in the lower corona rather than at the shock during the 
first stage, is again a likely indication of an accelerator that was distinct from the CME shock, and acted in addition to 
the shock, at lower altitude. This  is consistent with an interpretation of the electron acceleration in terms of 
reconnection in the corona behind the CME \citep{Trottet86,Cliver86,Kah:Hun-92}. New evidence for this interpretation has 
recently been provided by \cite{Aurass09} using UV and white light coronographic diagnostics along with radio data. 

%---------------------------------------------------------------------------------------
\subsection{CME interaction and related radio emission}
%---------------------------------------------------------------------------------------
Decametric-hectometric radio emission as a signature of CME interaction was discussed in some detail in two event studies 
\citep{Gopalswamy01,Gopalswamy02}. In both events the radio emission had a limited bandwidth, and was referred 
to as a continuum. In the present case this emission is likely a set of type~III bursts and was therefore labelled type~III$_3$. 
The starting frequency and the timing of these bursts are consistent with the idea that the electrons are accelerated  
while the faster following CME catches up with the slower preceding one. 

The association of CME interaction with particle acceleration has been ascribed by \citet{Gopalswamy01} to 
acceleration by the shock of the second CME as it traverses the previous one.
Problems of this interpretation were discussed by \cite{Kle-06}. 

{\bl Another important question is how } %KLK
the preceding CME could lead to a strengthening of the shock of the following one. This problem 
is still more evident in the 17 January 2005 event, because here the two CMEs are already extremely fast and 
likely to drive strong shocks even in the ambient solar wind. Their relative speed, however, is rather slow, so that efficient {\bl
acceleration by the shock of the second CME is not expected in the first CME. Another important feature is that the type III$_3$ 
bursts extend to the plasma frequency at the spacecraft. The electrons hence cannot propagate in closed magnetic structures related to the CMEs. 
The accelerator must release the electron beams onto open solar wind-type field lines. An alternative scenario to acceleration at the 
CME shock is again magnetic reconnection.}   
One can surmise that these rapid CMEs were preceded by 
sheath regions with strong magnetic fields of interplanetary origin, draped around the CME front. These regions 
are favourable for magnetic reconnection \citep[see the overview by][and references therein]{Forsyth06}. The 
high pressure in the sheath of the second CME will be further enhanced when its progression is slowed down by 
the previous CME. This makes the configuration favourable to magnetic reconnection involving  open solar 
wind field lines and strong magnetic fields, and allows one to understand qualitatively why accelerated electrons 
escape immediately towards the outer heliosphere.

In this scenario the type~III emission is expected to start close to the CME, in the upstream region. 
This can explain why the starting frequency is close to the frequency of the type~II burst, without implying that the electron beams 
were themselves accelerated at the shock.

%---------------------------------------------------------------------------------------
\subsection{Near-relativistic electrons at 1 AU}
%---------------------------------------------------------------------------------------
The flare/CME events under discussion were clearly  related with enhanced fluxes of near-relativistic 
electrons at  1 AU.  The peak intensity measured by ACE, of order $10^6 \; \rm (MeV \; cm^2 \; sr \;s)^{-1}$ 
in the 38-53 keV range, make the event comparable to 
the most intense ones of the sample studied by \cite{Haggerty02}, as seen in their Figure~3a. 
The CME speed is well above the speeds of the CMEs identified in that sample \citep{Simnett02}.

Since the energetic electrons were observed at 1 AU within the region exhibiting sheath-like characteristics following the shock on 17 January 2005, 
it is difficult to estimate electron travel times and to relate the {\it in situ} measurements to solar processes. 
But the observations strongly suggest that successive intensity increases are first due to the coronal acceleration 
in the flare/CME event, and then to an episode during the interaction of the two CMEs. The escape of these electrons 
to ACE confirms the view discussed above that the electrons cannot have been accelerated in the body of the first 
CME,  even if a shock driven by the second one passed through it. Neither can they originate from reconnection 
between closed magnetic field lines of the two CMEs. The electrons must rather be accelerated in regions from where 
they have ready access to solar wind magnetic field lines. This is consistent with a common acceleration of the 
mildly relativistic electrons and the electron beams at lower energies that produce the type III$_3$ emission.

%-------------------------------------------------------------------------------------------
\begin{acks}
This work was supported in part by the University of Athens Research Center (ELKE/EKPA).
 The authors appreciate discussions with and assistance of C. Caroubalos, C. Alissandrakis.
and S. Hoang. {\bl They would also like to thank an anonymous referee for 
many useful comments on the original manuscript} % KLK
\Ha~data were provided by the Kanzelh\"ohe Observatory, University
of Graz, Austria by M.~Temmer. 
The SOHO/LASCO data used here are produced by a consortium of the 
Naval Research Laboratory (USA), Max-Planck-Institut fuer Aeronomie (Germany), Laboratoire d'Astronomie 
(France), and the University of Birmingham (UK). The SoHO/LASCO CME catalog is generated and maintained 
at the CDAW Data Center by NASA and The Catholic University of America in cooperation with the Naval 
Research Laboratory. SOHO is a project of international cooperation between ESA and NASA.
KLK acknowledges the kind hospitality of the solar radio astronomy group at the University of Athens. 
\end{acks}
%-------------------------------------------------
%\bibliographystyle{spr-mp-sola-cnd}
%\bibliography{P01,P02,P03,P04,P05,P06,P07,SEE2007,Temp}

%-------------------------------------------------
\end{article}
\end{document}